\documentclass[11pt]{article}

\usepackage[a4paper,margin=1in]{geometry}
\usepackage[dvipsnames]{xcolor}
\usepackage{cite}
\usepackage{amsmath,amssymb,amsfonts}
\usepackage{algorithmic}
\usepackage{graphicx}
\usepackage{textcomp}
\usepackage{physics}
\usepackage{subcaption}
\usepackage{booktabs}
\usepackage{authblk}
\usepackage[hidelinks]{hyperref}

\title{\textbf{Hyperbolic Cluster States for Fault-Tolerant Measurement-Based Quantum Computing}}

\author[1, 2]{Ahmed Adel Mahmoud}
\author[3]{Gabrielle Tournaire}
\author[4]{Sven Bachmann}
\author[1, 2]{Steven Rayan}

\affil[1]{Centre for Quantum Topology and Its Applications (quanTA)} 
\affil[2]{Department of Mathematics and Statistics, University of Saskatchewan, Saskatoon, Canada}
\affil[3]{Department of Physics and Astronomy, The University of British Columbia, Vancouver, Canada}
\affil[4]{Department of Mathematics, The University of British Columbia, Vancouver, Canada}

\date{}

\begin{document}

\maketitle

\begin{center}
\texttt{ahmed.mahmoud@usask.ca} \quad
\texttt{gtournaire@phas.ubc.ca} \quad
\texttt{sbach@math.ubc.ca} \quad
\texttt{rayan@math.usask.ca}
\end{center}

\vspace{1em}

\begin{abstract}
Fault-tolerant measurement-based quantum computing (MBQC) provides a compelling framework for fault-tolerant quantum computation, in which quantum information is processed through single-qubit measurements on a three-dimensional entangled resource known as cluster state. To date, this resource has been predominantly studied on Euclidean lattices, most notably in the Raussendorf–Harrington–Goyal (RHG) construction, which underlies topological fault tolerance in MBQC.
In this work, we introduce the \emph{hyperbolic cluster state}, a generalization of the three-dimensional cluster state to negatively curved geometries, obtained via the foliation of periodic hyperbolic lattices. We present an explicit construction of hyperbolic cluster states and investigate their fault-tolerant properties under a realistic circuit-level depolarizing noise model. Using large-scale numerical simulations, we perform memory experiments to characterize their logical error rates and decoding performance.
Our results demonstrate that hyperbolic cluster states exhibit a fault-tolerance threshold comparable to that of the Euclidean RHG cluster state, while simultaneously supporting a constant encoding rate in the thermodynamic limit. This represents a substantial improvement in qubit overhead relative to conventional cluster-state constructions. These findings establish hyperbolic geometry as a powerful and experimentally relevant resource for scalable, fault-tolerant MBQC and open new avenues for leveraging negative curvature in quantum information processing.
\end{abstract}

\section{Introduction}

In recent years, physical systems defined on negatively curved spaces have attracted growing attention across several areas of condensed matter physics and quantum information science. This surge of interest has been driven in part by rapid progress in hyperbolic band theory \cite{maciejko2021hyperbolic, kienzle2022hyperbolic,maciejko2022automorphic, nagy2024hyperbolic}, the crystallography of hyperbolic lattices \cite{boettcher2022crystallography}, the discovery of hyperbolic topological phases \cite{wu2015topological, liu2022chern, liu2023higher, sun2024topological}, and the development of hyperbolic quantum error correction (QEC) codes \cite{albuquerque2009topological, breuckmann2016constructions, mahmoud2025systematic}. Collectively, these advances have established hyperbolic geometry as a powerful framework for realizing exotic physical phenomena that are inaccessible in flat, Euclidean settings.

Importantly, these theoretical developments have been accompanied by a growing number of experiments demonstrating that hyperbolic physics can be realized using existing laboratory platforms. By engineering effective negative curvature through tailored connectivity, experiments have implemented hyperbolic lattices using superconducting resonator networks, topo-electrical circuits, and related architectures \cite{kollar2019hyperbolic, lenggenhager2022simulating, huang2024hyperbolic, chen2023hyperbolic, zhang2022observation, zhang2023hyperbolic, dey2024simulating, bienias2022circuit, park2024scalable, xu2025scalable}. These results indicate that hyperbolic lattices are not merely theoretical constructs, but experimentally accessible structures with practical relevance.

Within QEC, topological codes defined on hyperbolic lattices provide a particularly compelling application of negative curvature. 
Unlike their Euclidean counterparts, such as the toric and surface codes, which suffer from a vanishing encoding rate in the thermodynamic limit, hyperbolic topological codes leverage a constant encoding rate while retaining locality \cite{kitaev2003fault, dennis2002topological, breuckmann2016constructions}. 
More recently, Floquet codes realized on hyperbolic lattices have been shown to reduce qubit overhead and improve performance relative to their original Euclidean honeycomb implementations \cite{higgott2024constructions, fahimniya2025fault}. These results suggest that hyperbolic geometry offers intrinsic advantages for fault-tolerant quantum information processing.
From a hardware perspective, hyperbolic QEC codes can be realized using a planar two-dimensional arrangement of physical qubits provided that the architecture supports sufficiently long-range couplings.
Such long-range interactions are available in several leading platforms, including trapped-ion processors, photonic architectures as well as cold atoms \cite{linke2017experimental, lekitsch2017blueprint, herrera2010photonic, larsen2019deterministic, maller2015rydberg, pichler2016measurement}.

Despite this progress, the role of hyperbolic lattices in MBQC has remained largely unexplored. The canonical three-dimensional RHG cluster state introduced by Raussendorf et al., which underpins fault-tolerant MBQC via topological error correction, was originally formulated on a Euclidean cubic lattice \cite{Raussendorf2006FaultTolerant, Raussendorf2007Topological}. However, the underlying construction relies only on local stabilizer structure and is therefore not inherently restricted to flat geometries. This observation naturally raises the question of whether hyperbolic lattices can serve as a superior substrate for MBQC.

In this work, we extend the three-dimensional RHG cluster-state construction to negatively curved spaces by introducing hyperbolic cluster states, obtained through the foliation of periodic hyperbolic $\{p,q\}$ lattices. We present an explicit realization of hyperbolic cluster states corresponding to a representative hyperbolic tessellation and systematically investigate its fault-tolerant properties. Using large-scale numerical simulations with a realistic circuit-level noise model, we perform memory experiments to assess the performance of the resulting MBQC scheme.

Our results demonstrate that hyperbolic lattices constitute a fundamentally advantageous foundation for MBQC. In particular, we show that hyperbolic cluster states exhibit an error threshold comparable to that of the RHG cluster state, while offering significantly higher encoding rates. This combination represents a remarkable improvement over Euclidean cluster-state constructions and highlights hyperbolic geometry as a promising direction for scalable, fault-tolerant MBQC.

\section{Hyperbolic Lattices}

A regular $\{p,q\}$ lattice is defined as a uniform tiling of a two-dimensional manifold by regular $p$-gons, such that exactly $q$ faces meet at every vertex. The geometry of the underlying space is determined by the Schläfli symbols $\{p,q\}$. In two-dimensional Euclidean space, regular tilings exist only when the angle-sum condition
\begin{equation}
    \frac{1}{p} + \frac{1}{q} = \frac{1}{2}
\end{equation}
is satisfied. This constraint admits only three solutions: the square lattice $\{4,4\}$, the hexagonal (honeycomb) lattice $\{6,3\}$, and its dual triangular lattice $\{3,6\}$. As a consequence, the space of regular Euclidean lattices is highly restricted.

By contrast, regular tilings of the hyperbolic plane satisfy the curvature constraint
\begin{equation}
    \frac{1}{p} + \frac{1}{q} < \frac{1}{2},
\end{equation}
leading to an infinite family of distinct $\{p,q\}$ lattices. These lattices inherit the intrinsic geometry of the hyperbolic plane, specifically its constant negative Gaussian curvature $K=-1$, which results in a number of vertices that grows exponentially with the graph radius.

For visualization and numerical construction, hyperbolic lattices are conveniently represented in the Poincaré disk model, which maps the hyperbolic plane to the unit disk conformally. In complex coordinates $z=x+iy$, the metric reads
\begin{equation}
    ds^2 = \frac{4\,|dz|^2}{(1 - |z|^2)^2},
\end{equation}
so that $ds^2=\lambda(z)^2\,(dx^2+dy^2)$ with conformal factor $\lambda(z)=2/(1-|z|^2)$. The conformal factor diverges as $|z|\to 1$, reflecting that the disk boundary lies at infinite hyperbolic distance. Consequently, hyperbolic geodesics are represented by circular arcs orthogonal to the boundary, and tiles that have equal hyperbolic area appear increasingly compressed in Euclidean size toward the edge of the disk.

\begin{figure}
    \centering
    \includegraphics[width=0.5\linewidth, trim= 8cm 8cm 6cm 6cm, clip]{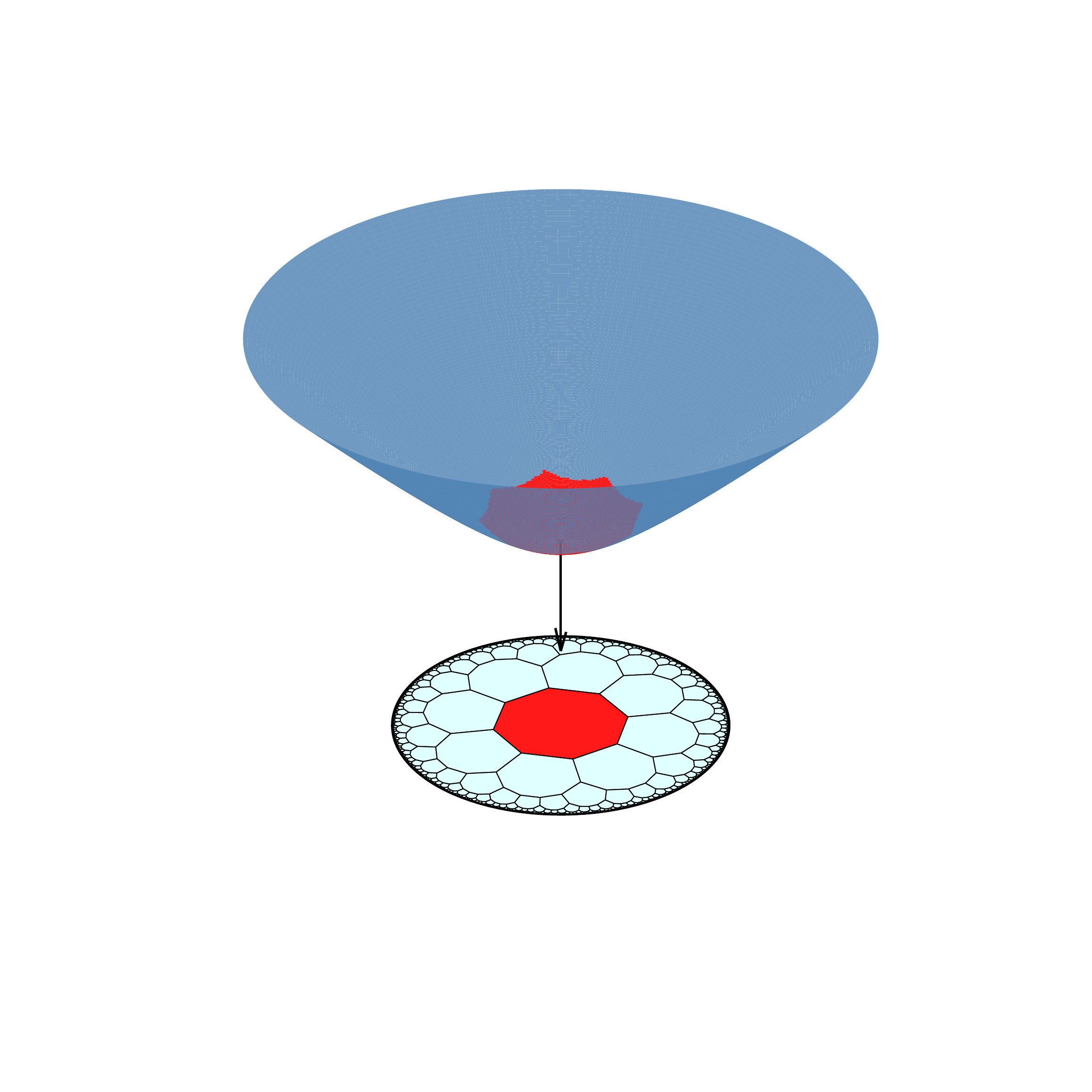}
    \caption{Stereographic projection of the regular hyperbolic $\{8,3\}$ lattice into the Poincaré disk with the unit cell highlighted in red.}
    \label{fig:projection}
\end{figure}

An example of the stereographic projection of the regular hyperbolic $\{8,3\}$ lattice into the Poincaré disk is shown in Fig.~\ref{fig:projection}. This lattice corresponds to the tessellation associated with the Bolza surface and serves as a canonical example throughout this work.

To obtain a finite lattice suitable for quantum information processing, periodic boundary conditions (PBCs) are imposed on a finite patch of the hyperbolic tiling. In this case, the resulting lattice with $F$ faces, $E$ edges, and $V$ vertices satisfies the combinatorial relation
\begin{equation}
    pF = 2E = qV,
    \label{F-E-V-relation}
\end{equation}
which follow by double-counting face--edge and vertex--edge incidences. Such a periodic lattice can be embedded into a closed two-dimensional Riemann surface $M$, whose topology is characterized by the Euler characteristic
\begin{equation}
    \chi(M) = F - E + V.
    \label{Euler-charac}
\end{equation}
When $\chi(M)$ is even, the surface $M$ is orientable and may be classified by its genus $g$ via
\begin{equation}
    \chi(M) = 2 - 2g.
    \label{surface=genus}
\end{equation}

A key distinction between Euclidean and hyperbolic lattices emerges at this topological level. Finite Euclidean lattices with PBCs necessarily satisfy $\chi(M)=0$, corresponding to a toroidal surface with genus $g=1$. In contrast, finite hyperbolic lattices have strictly negative Euler characteristic, $\chi(M) < 0$,
and therefore reside on surfaces of genus $g \geq 2$. In fact, Eqs.~\ref{F-E-V-relation}, \ref{Euler-charac}, and \ref{surface=genus} yield that
\begin{equation}
\label{eqn:genus-E-relation}
    g = 1 + E\left(\frac{1}{2}-\frac{1}{p}-\frac{1}{q}\right)
\end{equation}
showing that the genus grows with the `size' of the lattice. This intrinsic high-genus topology is central to the enhanced encoding properties of hyperbolic QEC codes and, as we demonstrate in this work, plays a crucial role in the structure and performance of hyperbolic cluster states for MBQC.

Another key distinction is the nature of the lattice symmetry group. For Euclidean lattices, the translation symmetry group is Abelian (e.g., $\mathbb{Z}\times\mathbb{Z}$ for the square lattice); in contrast, hyperbolic lattices are generated by non-Abelian discrete isometry groups, commonly referred to as Fuchsian groups. A \emph{Fuchsian translation group} $\Gamma$ is a torsion-free Fuchsian group that admits a finite presentation of the form
\begin{equation}
\Gamma = \langle \gamma_1,\ldots,\gamma_{p_B/2} \mid X_{{p_B,q_B}}=\mathbb{I}\rangle,
\end{equation}
where $\{p_B,q_B\}$ denotes the Bravais lattice associated with the $\{p,q\}$ tiling and $X_{{p_B,q_B}}$ encodes the defining group relations \cite{boettcher2022crystallography}. Realizing a finite periodic sublattice of a regular $\{p,q\}$ hyperbolic tiling corresponds to choosing a finite-index normal subgroup $\Gamma_{\mathrm{PBC}}\subset \Gamma$. Computational group-theory algorithms enable the enumeration of finite-index normal subgroups of finitely presented groups, which can be used to construct such periodic lattices. In this work, we perform the enumeration process using the computational algebra system GAP \cite{GAP4}; specifically, we use the LINS package, which implements the low-index normal subgroup algorithm \cite{LINS0.9, dietze1974determining}. The full lattice construction procedure is outlined in \cite{mahmoud2025systematic}.

\section{Hyperbolic QEC codes}
\label{sec:hyperbolic_qec}
Hyperbolic QEC codes form a family of Calderbank–Shor–Steane (CSS) stabilizer codes obtained by generalizing Kitaev’s toric code from Euclidean to negatively curved lattices \cite{albuquerque2009topological, breuckmann2016constructions}. These codes are defined on regular hyperbolic $\{p,q\}$ lattices with PBCs and inherit both their stabilizer structure and encoding properties from the underlying geometry. In particular, physical qubits are placed on the edges of the lattice. Moreover, stabilizer generators are associated with faces and vertices: for each face $f$, a plaquette stabilizer
\[
S_Z(f) = \prod_{e \in \partial f} Z_e
\]
acts by Pauli-$Z$ on the qubits along the boundary of $f$, while for each vertex $v$, a site stabilizer
\[
S_X(v) = \prod_{e \ni v} X_e
\]
acts by Pauli-$X$ on all edges incident to $v$. By construction, all stabilizers commute, and the resulting code corrects both bit-flip and phase-flip errors in the CSS framework.

A defining feature of hyperbolic QEC codes is their nonvanishing encoding rate. For a finite hyperbolic lattice with $n=E$ physical qubits and PBCs, the number of logical qubits is determined by the genus of the associated closed surface as $k = 2g$. By Eq.~\ref{eqn:genus-E-relation}, the resulting encoding rate, namely the ratio of the number of encoded qubits over the number of physical qubits, is
\begin{equation}
\label{eqn:encoding_rate}
    \frac{k}{n} = 1 - \frac{2}{p} - \frac{2}{q} + \frac{2}{n},
\end{equation}
which approaches a constant value in the thermodynamic limit $n \to \infty$. This behavior stands in sharp contrast to the surface and toric codes defined on Euclidean lattices, whose encoding rates vanish asymptotically.

From Eq.~\ref{eqn:encoding_rate}, the encoding rate of a hyperbolic QEC code can be increased by choosing larger values of $p$ or $q$, corresponding to higher-weight plaquette or site stabilizers, respectively. This trade-off reflects a fundamental tension between encoding rate and local stabilizer weight: increasing stabilizer weight improves the global encoding efficiency but degrades local error detectability and typically reduces the error threshold.

As is the case in the toric code \cite{kitaev2003fault,Bachmann2017LocalDisorder}, logical operators in hyperbolic QEC codes are naturally characterized in homological terms. $Z$-type logical operators correspond to nontrivial cycles, while $X$-type logical operators correspond to nontrivial cocycles in the lattice graph. In this work, we identify all plaquette boundaries and logical operators using the Hyperbolic Cycle Basis algorithm introduced in \cite{mahmoud2025systematic}, which provides an efficient and systematic method for computing canonical cycle representatives on finite hyperbolic lattices.

Despite their favorable encoding rates, hyperbolic CSS codes exhibit relatively short code distances compared to the toric code, a consequence of the encoding rate-distance tradeoff 

$$kd^2 \leq c (\log{k})^2 n,$$
where $c$ is a constant \cite{delfosse2013tradeoffs}. Specifically, the distance $d$, which is the minimum length of a nontrivial cycle, scales only logarithmically with the number of physical qubits, $d = \mathcal{O}(\log n)$, in contrast to the square-root scaling, $d = \mathcal{O}(\sqrt{n})$, typical of Euclidean surface codes. As a result, the asymptotic code parameters of a hyperbolic QEC code are given by
\begin{equation}
     [[n,\,(1 - \tfrac{2}{p} - \tfrac{2}{q})n + 2,\;\mathcal{O}(\log n)]].
\end{equation}

\section{Foliated CSS QEC Codes}
\label{sec:foliated_css}

Measurement-based quantum computing was introduced in the one-way model of Raussendorf and Briegel \cite{Raussendorf2001OneWay}. In this paradigm, a highly entangled cluster state is prepared on a graph, and computation is enacted by single-qubit measurements whose outcomes determine classical feed-forward corrections. A one-dimensional cluster state already enables measurement-based teleportation of quantum information along the chain, while a two-dimensional cluster state provides a universal resource for quantum computation \cite{Raussendorf2003MeasurementBased}. 

Fault tolerance, however, requires active error correction. Topological MBQC addresses this by embedding quantum error-correcting structure directly into the measurement pattern of a three-dimensional cluster state. In the RHG construction, the three-dimensional cluster state realizes a foliated version of a two-dimensional surface code, and local measurement outcomes combine into parity checks whose syndromes diagnose faults \cite{Raussendorf2006FaultTolerant, Raussendorf2007Topological, Raussendorf_2005}. Logical protection is then controlled by the code distance and by the separation of the corresponding topological structures in the three-dimensional lattice, yielding a finite threshold for local noise. More generally, the RHG construction can be extended beyond the surface code: the notion of \emph{foliation} provides a systematic route to construct fault-tolerant cluster-state resources from arbitrary CSS stabilizer codes \cite{Bolt2016FoliatedQECC}. 

A primal foliated cluster-state layer associated with a CSS code is constructed by introducing qubits for each data qubit of the code and additional ancilla qubits corresponding to its stabilizer checks. Concretely, one prepares all data qubits in the $\ket{+}$ state, and prepares an ancilla qubit in the $\ket{+}$ state for each $Z$-type stabilizer generator. One then applies \textsc{CZ} gates between a $Z$-check ancilla and every data qubit in the support of that stabilizer, producing a bipartite graph state that encodes the $Z$-check structure in its adjacency. For topological CSS codes defined on a $\{p,q\}$ tessellation, each ancilla associated with a $Z$-type check couples to the $p$ data qubits along the boundary of the corresponding face; we refer to these qubits as \emph{face ancillas} to distinguish them from the \emph{node ancillas} associated with $X$-type checks.

Since for each face ancilla $f$ and adjacent data qubit $e$ we have
\[
CZ_{f,e}\, X_f = X_f Z_e\, CZ_{f,e},
\]
and the entangling circuit applied to the planar code results in a state on the layer that is stabilized by
\begin{equation}\label{eq:cluster stabilizer primal layer}
  K_f \;=\; X_f \prod_{e\in \partial f} Z_e  \quad \mathrm{and} \quad  K_e= X_e \prod_{f:e\in \partial f} Z_f.
\end{equation}

Measuring the ancilla $f$ in the $X$ basis with outcome $m_f\in\{\pm1\}$ projects the remaining (unmeasured) qubits onto an eigenstate of the plaquette operator
\[
S_Z(f)=\prod_{e\in \partial f} Z_e
\]
with eigenvalue $m_f$. In this way, the $Z$-type stabilizer syndrome is obtained directly from the ancilla measurement outcomes.

The entangled state constitutes a single (two-dimensional) cluster-state layer associated with the $Z$-check structure of the CSS code. A three-dimensional foliated cluster state is obtained by alternating such layers with layers associated with the complementary $X$-check structure. For topological CSS codes, this is naturally described in terms of the dual cellulation: the dual code is obtained by exchanging the roles of faces and vertices (equivalently, of $Z$- and $X$-type checks), so that $Z$-check (face) ancillas in a primal layer are replaced by $X$-check (node) ancillas in the adjacent dual layer. The two layers are then coupled by applying a \textsc{CZ} gate between each data qubit in the primal layer and its corresponding data qubit in the dual layer. Repeating this alternating construction for $2z$ layers yields the full foliated cluster state. For self-dual tilings (e.g., the square $\{4,4\}$ tiling underlying the surface code), the primal and dual layers have the same local structure, although they still correspond to distinct check types.

The resulting foliated resource can be described equivalently on a three-dimensional complex. Data qubits in the primal layer and node ancillas correspond to edges of this complex, while face ancillas and dual data qubits correspond to faces. In particular, in a primal layer the face ancillas are adjacent to the data qubits on the corresponding plaquette boundaries. In a dual layer a node ancilla corresponds to the vertical edge shared by the $q$ faces corresponding to the adjacent dual data qubits. The cluster state is then the unique common $+1$ eigenstate of similar stabilizers as in \eqref{eq:cluster stabilizer primal layer}
for all faces $f$ and edges $e$ in the three-dimensional complex. Note that in the dual complex, a primal edge $e$ is mapped to a dual face $\overline{f}$ and vise-versa, yielding the stabilizers
\begin{equation}\label{eq:cluster stabilizers 3D}
K_f = X_f \prod_{e\in \partial f} Z_e, \quad \mathrm{and} \quad  K_{\overline{f}}=X_{\overline f} \prod_{\overline e\in \partial \overline f} Z_{\overline e}.
\end{equation}

Fault-tolerant MBQC and syndrome extraction arise from local measurement patterns and the induced parity constraints on measurement outcomes. As shown in \cite{Bolt2016FoliatedQECC}, for a foliated CSS construction one can measure all qubits except the data qubits on the final layer in the $X$ basis; the resulting measurement pattern teleports the logical state encoded in the initial layer to the final layer, up to known Pauli byproduct operators determined by the outcomes. Operationally, the foliated direction plays the role of a discrete time coordinate, and intermediate measurement outcomes provide syndrome information that can be used for error correction \cite{Raussendorf_2005}.

A convenient way to express the resulting parity checks is through products of graph-state stabilizers over closed surfaces in the associated three-dimensional complex. 
Consider the cluster state stabilizers \eqref{eq:cluster stabilizers 3D}. Let $M$ be a closed surface in the primal complex, then the product of stabilizers over the faces contained in $M$ yields
\begin{equation}
\prod_{f\in M} K_f
\;=\;
\left(\prod_{f\in M} X_f\right)
\left(\prod_{e\in \partial M} Z_e\right)
\;=\;
\prod_{f\in M} X_f,
\end{equation}
where the $Z$ operators cancel because each interior edge contributes twice and $\partial M=0$. Consequently, in the absence of errors, the product of the corresponding single-qubit $X$-measurement outcomes is constrained to be $+1$. A $Z$-type fault on any measured qubit in $M$ flips its $X$-measurement outcome and therefore flips the parity check, producing a syndrome defect. An analogous family of checks arises from closed surfaces on the dual complex, yielding complementary syndrome information. 

Finally, we emphasize that the ability to extract local parity checks from measurement outcomes, and hence to perform fault-tolerant decoding with a nonzero noise threshold, is not unique to cluster states obtained by foliation. For example, \cite{nickerson2018measurementbasedfaulttolerance} presents three-dimensional cluster-state resources that support fault-tolerant MBQC yet are not constructed by foliating a stabilizer code, and \cite{Newman2020generatingfault} proposes systematic procedures for generating fault-tolerant cluster states from crystalline structures and analyzes their performance under circuit-level noise.

\section{The Hyperbolic Cluster State}
\label{sec:HCS}
We now augment the foliated CSS construction of Sec.\ref{sec:foliated_css} by instantiating it with the hyperbolic QEC codes introduced in Sec.\ref{sec:hyperbolic_qec}.
Starting from a periodic $\{p,q\}$ hyperbolic code with data qubits placed on edges and stabilizer generators defined on faces and vertices, we build a three-dimensional cluster-state resource by stacking $2z$ layers and coupling them along a discrete foliated direction. As in the general foliated construction, the layers alternate between \emph{primal} and \emph{dual} sheets: primal layers contain the face ancillas associated with $Z$-type checks, while dual layers contain the node ancillas associated with $X$-type checks. Each layer is thus a bipartite Tanner graph encoding the incidence relations between data qubits and the corresponding stabilizer generators, and inter-layer \textsc{CZ} couplings connect each data qubit to its counterparts in the adjacent layers.

\begin{figure*}[t]
    \centering
    \begin{subfigure}[t]{0.49\linewidth}
        \centering
        \includegraphics[width=\linewidth, trim= 1cm 1cm 0cm 0cm, clip]{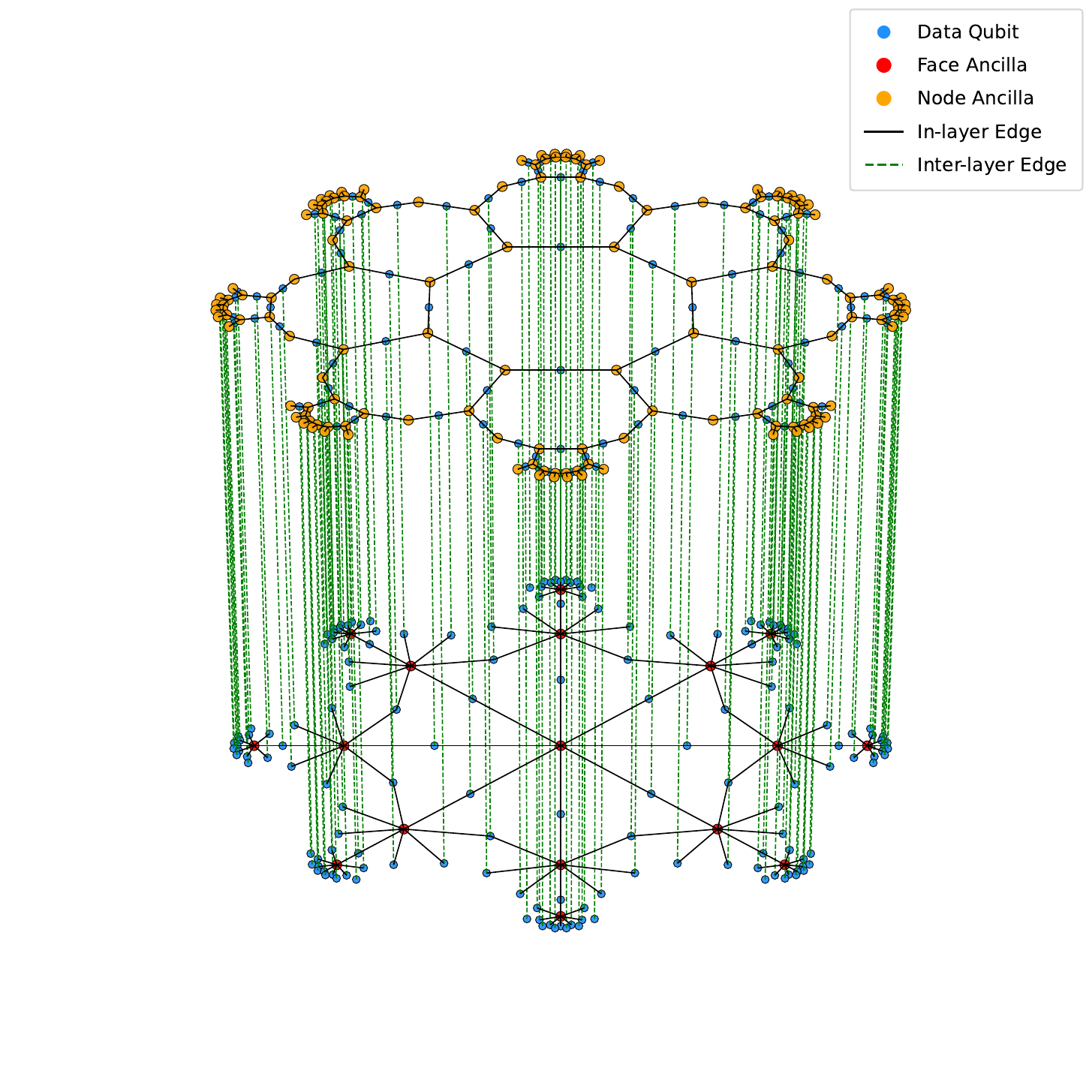}
    \caption{Two consecutive layers of a foliated hyperbolic cluster state constructed from the $\{8,3\}$ hyperbolic CSS code. The first and second layers correspond to the primal and dual Tanner graphs, respectively, and each edge represents an entangling \textsc{CZ} operation in the cluster-state preparation circuit.}
        \label{fig:foliated_cs_hyperbolic}
    \end{subfigure}
    \hfill
    \begin{subfigure}[t]{0.49\linewidth}
        \centering
        \includegraphics[width=\linewidth, trim= 0cm 2cm 0cm 0cm, clip]{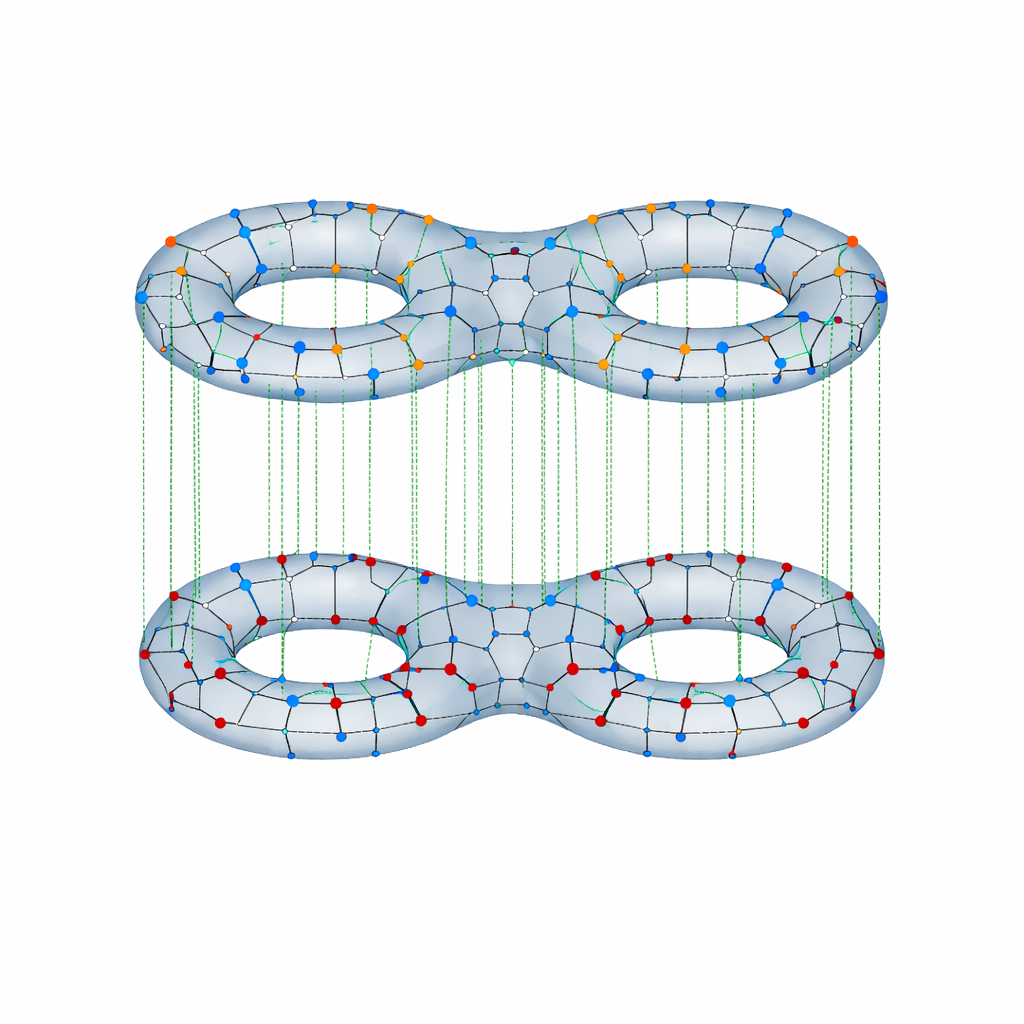}
        \caption{Schematic of the foliated hyperbolic cluster state after imposing periodic boundary conditions. Each layer is embedded into a closed higher-genus surface (genus $g=2$ shown for illustration), while dashed edges indicate inter-layer couplings that generate the full three-dimensional cluster-state connectivity.}
        \label{fig:foliated_cs_tori}
    \end{subfigure}

    \caption{Two-layer foliated hyperbolic cluster state derived from an $\{8,3\}$ lattice before and after imposing PBCs.}
    \label{fig:foliated_cs}
\end{figure*}

Figure~\ref{fig:foliated_cs} shows an explicit instance of this construction for the $\{8,3\}$ hyperbolic QEC code. For visual clarity, only two consecutive layers (one primal and one dual) are displayed, where edges indicate the \textsc{CZ} couplings used in state preparation. After imposing PBCs, each layer is embedded on a closed surface of genus $g\ge 2$, and the full resource forms a three-dimensional complex built from these higher-genus sheets coupled along the foliated direction. Note that for the $\{8,3\}$ instance shown in Fig.~\ref{fig:foliated_cs_hyperbolic}, imposing PBCs yields an embedding on a closed surface of genus $g=10$. In Fig.~\ref{fig:foliated_cs_tori} we instead depict a genus-$2$ surface as a schematic visualization of the periodic identification.

\begin{figure*}[t]
    \centering
    \begin{subfigure}[t]{0.49\linewidth}
        \centering
        \includegraphics[width=\linewidth, trim=0cm 0cm 0cm 3.5cm, clip]{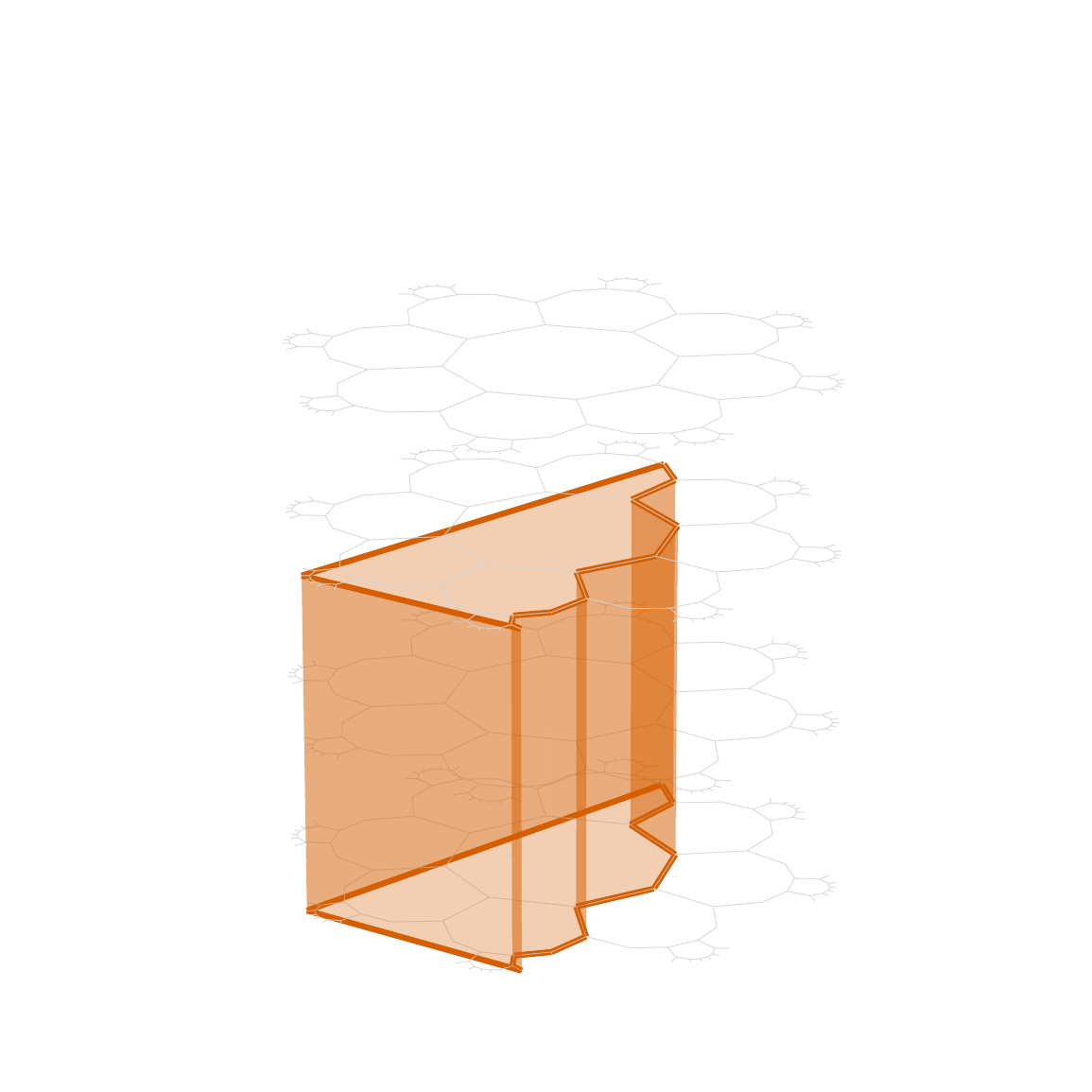}
    \caption{$Z$-type correlation surface obtained by lifting a nontrivial primal 1-cycle of the hyperbolic CSS code into the foliated cluster-state geometry. Owing to the stabilizer structure, the correlation surface is supported exclusively on primal layers, here shown for $z\in\{0,2\}$.}
        \label{fig:Z_corr_surfaces}
    \end{subfigure}
    \hfill
    \begin{subfigure}[t]{0.49\linewidth}
        \centering
        \includegraphics[width=\linewidth, trim=0cm 0cm 0cm 3.5cm, clip]{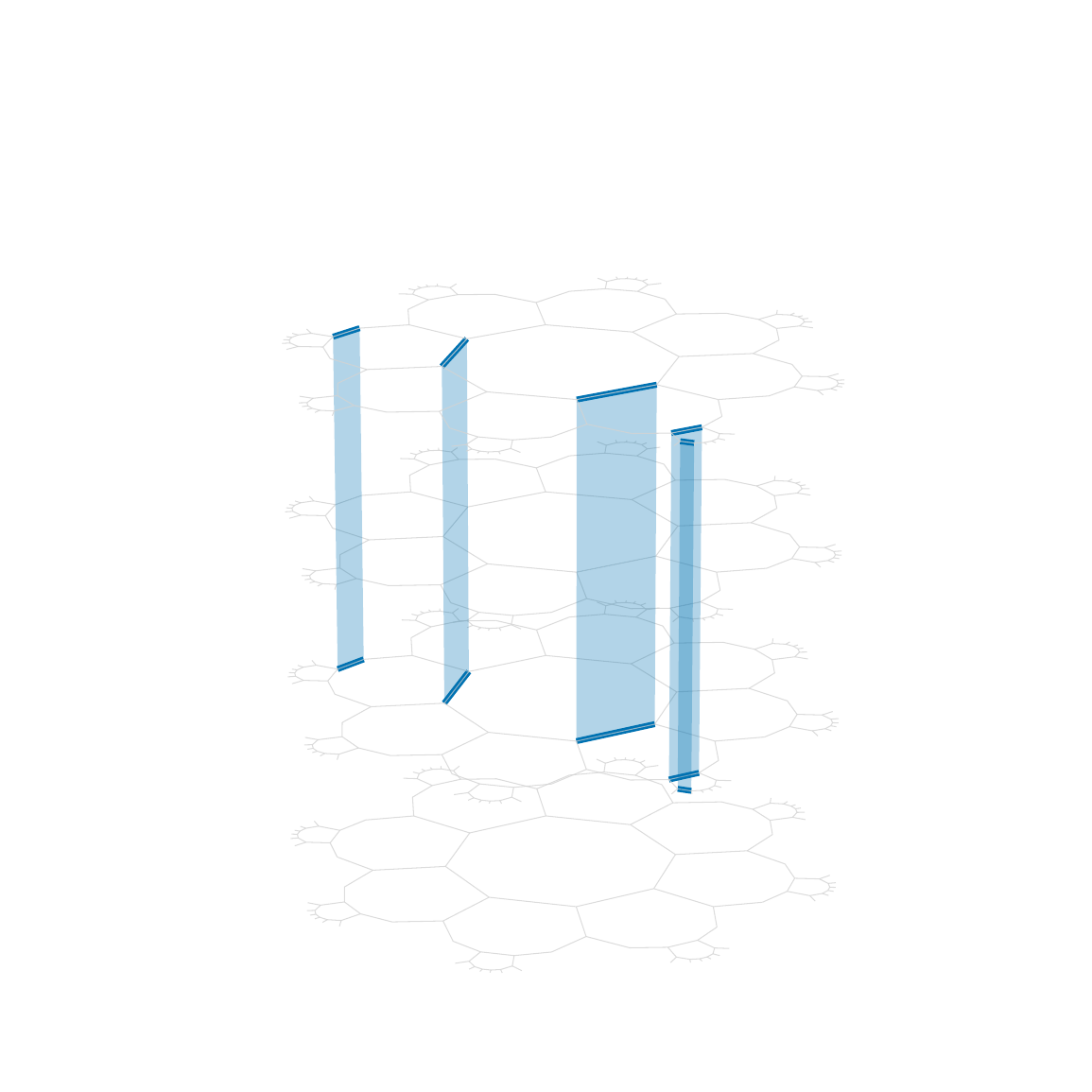}
    \caption{$X$-type correlation surface obtained by lifting a nontrivial dual 1-cocycle into the foliated cluster-state geometry. The surface is supported exclusively on dual layers, here shown for $z\in\{1,3\}$.}
        \label{fig:X_corr_surfaces}
    \end{subfigure}

\caption{Representative correlation surfaces of the hyperbolic cluster state obtained by foliating the $\{8,3\}$ hyperbolic QEC code. These surfaces realize the logical degrees of freedom of the underlying QEC code.}
    \label{fig:corr_surfaces}
\end{figure*}

In this periodic setting, the nontrivial topology of the underlying surface lifts directly to the cluster-state resource and determines its logical structure. As reviewed in Sec.\ref{sec:hyperbolic_qec}, logical operators of the hyperbolic CSS code correspond to nontrivial homology and cohomology classes: $Z$-type logical operators are represented by nontrivial 1-cycles (elements of $H_{1}$), while $X$-type logical operators are represented by nontrivial 1-cocycles (elements of $H^{1}$, equivalently 1-cycles on the dual complex). Upon foliation, a representative 1-cycle $\gamma$ is promoted to a two-dimensional object by sweeping it along the foliated direction $z$, yielding a 2-chain $\Sigma$ in the three-dimensional cell complex. In this periodic setting, while the temporal boundaries persist, the spatial boundary contributions cancel so that $\partial\Sigma=0$, and $\Sigma$ defines a \emph{correlation surface}. Concretely, a $Z$-type correlation surface is obtained by lifting a 1-cycle in this manner, while an $X$-type correlation surface is obtained by lifting a 1-cocycle. Because primal and dual layers support distinct parity checks, $Z$-type correlation surfaces are supported on primal layers, whereas $X$-type correlation surfaces are supported on dual layers as shown in Fig.~\ref{fig:Z_corr_surfaces} and Fig.~\ref{fig:X_corr_surfaces}, respectively.

\begin{figure*}[t]
    \centering
    \begin{subfigure}[t]{0.48\linewidth}
        \centering
        \includegraphics[width=\linewidth, trim=2.5cm 3cm 0cm 0cm, clip]{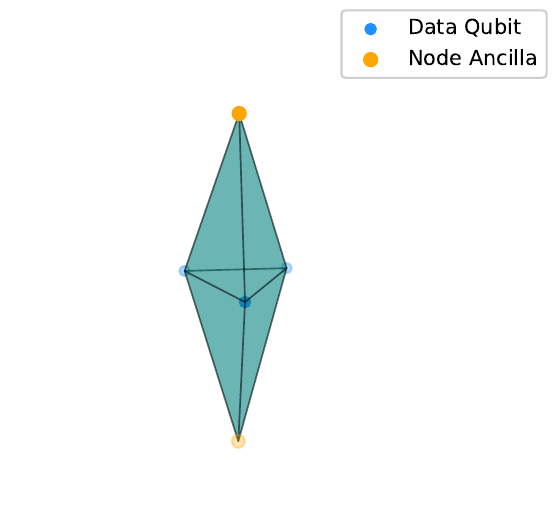}
    \end{subfigure}
    \hfill
    \begin{subfigure}[t]{0.48\linewidth}
        \centering
        \includegraphics[width=\linewidth, trim=2.5cm 3cm 0cm 0cm, clip]{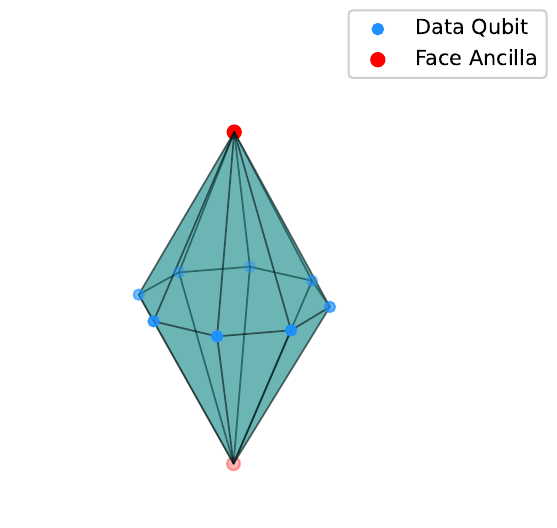}
    \end{subfigure}
   \caption{Geometric realization of parity-check operators in the hyperbolic cluster state obtained from the $\{8,3\}$ hyperbolic QEC code. The left (right) panel shows an $X$-type ($Z$-type) parity check, supported on a three-dimensional bi-pyramidal structure whose equatorial face lies in a single layer and whose apices extend to the neighboring layers. The parity-check measurement outcome is given by the sum (mod 2) of the single-qubit measurement outcomes associated with the qubits represented by the nodes of the corresponding structure.}
    \label{fig:parity-checks}
\end{figure*}

Finally, the parity checks extracted from single-qubit measurements admit a simple geometric interpretation in the foliated hyperbolic setting. Each check is supported on a three-dimensional \emph{bi-pyramidal} neighborhood whose equator lies in a single layer and whose two apices lie in the adjacent layers above and below. The base of the bi-pyramid is inherited from the local tiling geometry and therefore reflects the Schläfli symbol $\{p,q\}$. For the $\{8,3\}$ hyperbolic cluster state, the resulting $X$-type checks correspond to bi-pyramids with triangular equators (set by $q=3$), while the $Z$-type checks correspond to bi-pyramids with octagonal equators (set by $p=8$), as illustrated in Fig.~\ref{fig:parity-checks}. Together, the correlation surfaces and bi-pyramidal check operators provide a geometric characterization of the logical and stabilizer degrees of freedom relevant for fault-tolerant MBQC on hyperbolic cluster states.

\section{Quantum Error Correction on the Hyperbolic Cluster State}
\label{section:QEC on HCB}

Having established the construction protocol of the hyperbolic cluster state, its parity check operators and correlation surfaces, we now describe the noise model and decoding pipeline used to perform memory experiments and estimate fault-tolerance thresholds.

\subsection{Memory-Experiment Protocol and Scheduling}

A memory experiment probes the ability of the hyperbolic cluster-state construction to preserve logical quantum information across the measurement sequence (equivalently, to teleport an encoded state from the input layer to the output layer) in the presence of noise. Operationally, it consists of preparing the entangled cluster state, measuring all qubits, extracting syndromes from the measurement outcomes, and decoding to infer corrections. Concretely:

\begin{enumerate}
    \item \textbf{Initialization:} prepare every qubit (data and ancilla) in the state $\ket{+}$.
    \item \textbf{Entangling circuit:} apply \textsc{CZ} gates according to the cluster-state graph. Within each layer, \textsc{CZ} gates between a check ancilla and its neighboring data qubits are applied in a fixed local order around each check (counterclockwise in this work, see Fig.~\ref{fig:CZ_schedule}); inter-layer \textsc{CZ} gates then couple corresponding data qubits in adjacent layers.
    \item \textbf{Readout:} measure every qubit in the $X$ basis. Parity checks are computed as predetermined products of the single-qubit outcomes.
\end{enumerate}
The role of the entangling-gate schedule is to control \emph{fault propagation}: a locally consistent ordering ensures that single faults propagate into predominantly string-like error patterns, which enables efficient decoding via minimum-weight perfect matching (MWPM). We use the same schedule throughout all simulations.

\subsection{Circuit-Level Noise Model}

We adopt a standard circuit-level depolarizing noise model. After each \textsc{CZ} gate, an independent two-qubit Pauli error
\[
E \in \{I,X,Y,Z\}^{\otimes 2}\setminus\{I\otimes I\}
\]
is applied with total probability $p$ (uniformly over the $15$ nontrivial Paulis, i.e., each occurs with probability $p/15$). Each single-qubit measurement is preceded by an independent Pauli-$X$ (bit flip) error with probability $p$.

Let the underlying periodic $\{p,q\}$ hyperbolic lattice have $F$ faces, $E$ edges, and $V$ vertices, and let the cluster state consist of $2z$ layers (i.e., $z$ primal and $z$ dual layers). The total number of \textsc{CZ} gates equals the number of edges of the cluster-state graph; decomposing these into intra- and inter-layer couplings gives
\begin{equation}
\label{eq:CZcount}
    CZ_{\mathrm{HCS}} = z(pF + qV) + E(2z-1) 
    = E(6z - 1),
\end{equation}
where we used Eq.~\ref{F-E-V-relation} to suppress $F$ and $V$ in favor of $E$. The total number of single-qubit measurements equals the number of vertices in the cluster-state graph
\begin{equation}
\label{eq:Mcount}
      M_{\mathrm{HCS}} = z(F + V + 2E)
    = 2Ez \left(1 + \frac{1}{p} + \frac{1}{q}\right).
\end{equation}
Under our noise model, each \textsc{CZ} location contributes $15$ distinct Pauli faults, and each measurement contributes one flip fault, so the total number of distinct single-location fault processes is
\begin{equation}
\label{eq:NEcount}
\begin{aligned}
    N_F &= 15\, CZ_{\mathrm{HCS}} + M_{\mathrm{HCS}} \\
        &= 2Ez\left(46 + \frac{1}{p} + \frac{1}{q} - \frac{15}{2z}\right).
\end{aligned}
\end{equation}

\subsection{Syndrome Extraction and an Effective Noise Model}

To incorporate circuit-level fault propagation into decoding, we first characterize the syndrome produced by each single fault location. Specifically, we simulate all $N_F$ single-fault circuits obtained by inserting exactly one fault at a specified location, and then running the remainder of the ideal circuit. Each such simulation yields a deterministic syndrome pattern (a set of violated parity checks) associated with that fault. 

To illustrate the large scale of the performed simulations, consider the $\{8,3\}$ hyperbolic cluster state with $E=600$ edges and $2z=8$ foliated layer pairs, simulated below. The total number of distinct single-fault circuits for this hyperbolic cluster state is $N_F=2.14\times10^{5}$, each simulated independently. 

Many single faults lead to identical syndromes, while some lead to the trivial syndrome. We therefore group faults by their resulting syndrome and aggregate their probabilities. This grouping defines an \emph{effective} description at the level of data-qubit error events: for each data qubit, we compute an effective error probability $P_d$ by summing the probabilities of all single-location faults whose propagated effect includes that data qubit in the induced error chain, with the appropriate Pauli types treated separately in the CSS setting.

\begin{figure}
    \centering
    \includegraphics[width=0.5\linewidth]{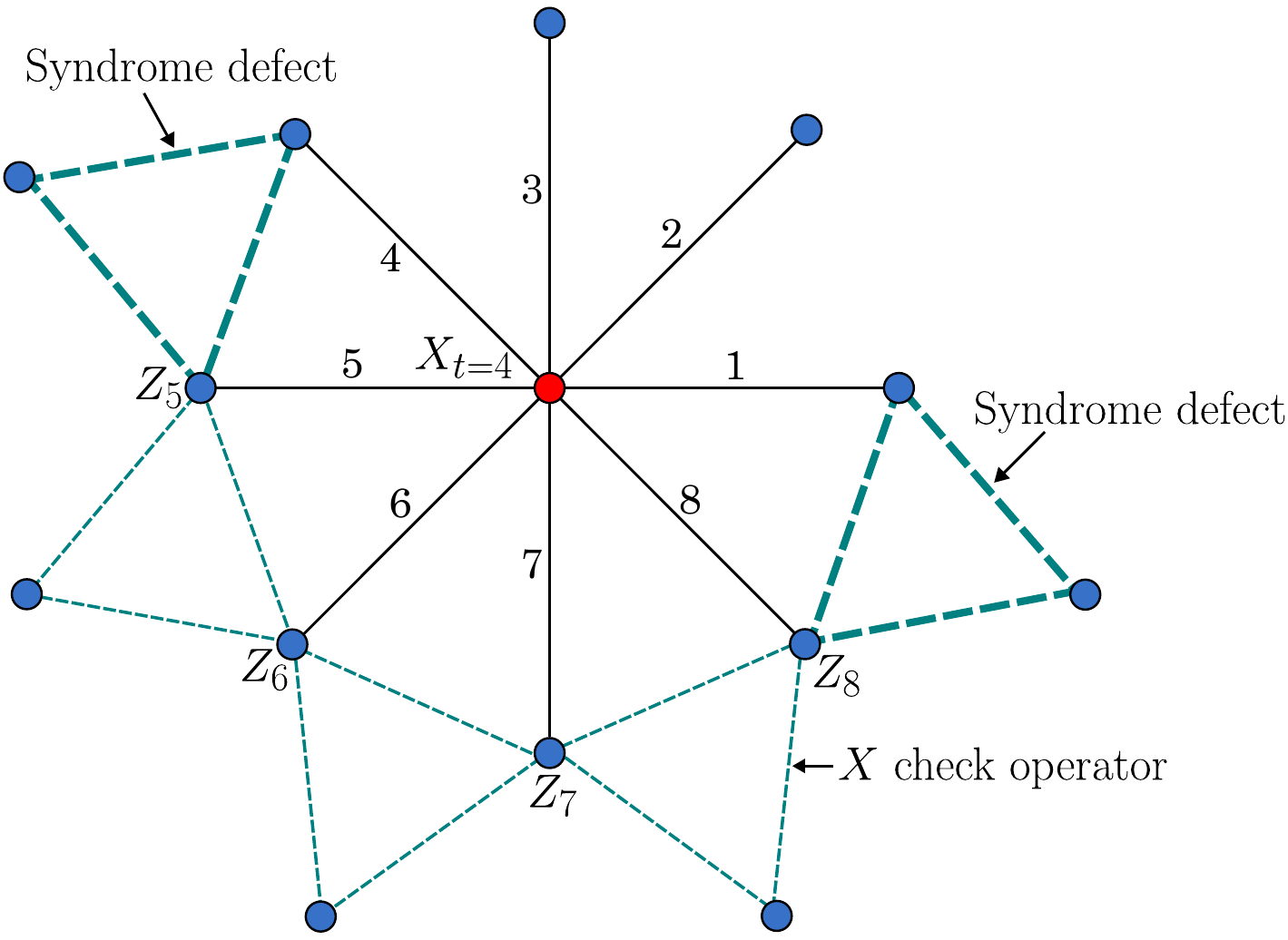}
    \caption{\textsc{CZ}-gate schedule around a primal plaquette, applied in a fixed counterclockwise order. An $X$ fault on the face ancilla occurring after the fourth \textsc{CZ} gate propagates through the remaining entangling operations, inducing a contiguous chain of $Z$ faults on the subsequent neighboring data qubits. This chain flips the outcomes of the two $X$-type parity checks at its endpoints, producing a correlated pair of syndrome defects (thick blue dashed triangles). The other $X$-type check operators involved are represented by thinner blue dashed triangles. Since they contain two $Z$ errors each, they do not produce a syndrome deffect. }
    \label{fig:CZ_schedule}
\end{figure}

Because \textsc{CZ} is a Clifford gate, Pauli faults propagate predictably. In particular, an $X$ fault on an ancilla can propagate to neighboring data qubits as a string of $Z$ faults, producing a pair of syndrome defects at the endpoints of that string. The fixed local scheduling around each check is chosen so that such propagated errors remain contiguous along the check neighborhood. Figure~\ref{fig:CZ_schedule} illustrates a representative mechanism: an $X$ error on the ancilla occurring after the fourth \textsc{CZ} gate can be commuted through the remaining \textsc{CZ} gates as
\begin{equation}
\label{eq:ancillaXprop}
    \Big(\prod_{q=5}^{8} CZ_{a,q}\Big)\, X_a
    \;=\;
    X_a\Big(\prod_{q=5}^{8} Z_q\Big)\Big(\prod_{q=5}^{8} CZ_{a,q}\Big),
\end{equation}
so that the ancilla fault induces a contiguous $Z$-error chain on the neighboring data qubits. The resulting syndrome is detected on the parity checks at the chain endpoints (highlighted in Fig.~\ref{fig:CZ_schedule}). Note that this correlated error is equivalent to $Z$ errors on code qubits 1, 2, 3 and 4, up to a plaquette stabilizer. More generally, single faults can yield correlated defect pairs separated by distances set by the local face size, with maximum separation on the order of $p/2$ along a face boundary.

The effective probabilities $\{P_d\}$ on each edge $e$ are converted into edge weights on the decoding graph via
\[
W(e) = -\ln P_d(e),
\]
so that MWPM  returns the inferred error under the weighted (effective) decoding graph. We construct separate weighted decoding graphs for $Z$- and $X$-type decoding, corresponding to the primal and dual layer structures of the foliated complex.

\subsection{Decoding and Logical-Error Identification}

To estimate logical error rates, we perform Monte Carlo simulations in which faults are sampled according to the circuit-level noise model and applied to the full cluster-state circuit. The measured $X$-basis outcomes are processed into a syndrome (a set of violated parity checks), and we decode this syndrome using MWPM on the corresponding weighted decoding graph. The MWPM output is a set of matched defect pairs.

To convert each matched pair into an explicit correction, we work on a separate unweighted connectivity graph that encodes only the geometry of admissible error strings. Concretely, this graph is the stack of the underlying periodic lattice across the relevant layers: for $Z$ decoding we use the primal stack, while for $X$ decoding we use the dual stack. Vertices of this connectivity graph represent possible defect locations, and edges represent elementary local steps by which an error string can propagate between neighboring checks. For each matched pair, we therefore select a correction chain as a shortest path between the two defects on this connectivity graph. Taking the sum (mod~2) of the paths over all matched pairs yields the inferred correction set.

It is convenient to represent both actual faults and inferred corrections at the level of supports as $\mathbb{Z}_2$ 1-chains on data qubits (i.e., subsets of data qubits modulo symmetric difference. Let $E_{\mathrm{inf}}$ denote the inferred correction chain obtained by taking the mod-$2$ sum of the shortest-path chains over all matched pairs, and let $E_{\mathrm{act}}$ denote the corresponding effective data-qubit error chain induced by the sampled circuit faults. The residual chain is then
\[
E_{\mathrm{res}} = E_{\mathrm{act}} + E_{\mathrm{inf}} \quad (\mathrm{mod}\;2).
\]
When translated back to Pauli operators, $E_{\mathrm{res}}$ corresponds to the residual Pauli error (e.g., $Z(E_{\mathrm{res}})$ in the $Z$-decoding problem). A logical failure is declared if $E_{\mathrm{res}}$ has nontrivial pairing with at least one logical correlation surface, equivalently if the corresponding residual Pauli operator anticommutes with a logical operator defined by that surface. Because the underlying code is CSS, $Z$- and $X$-type decoding are performed independently using the primal and dual decoding graphs, respectively.

\section{Fault-Tolerant Computation on Hyperbolic Cluster States}

In this section, we outline two complementary paradigms for fault-tolerant computation that inform the development of logical operations on hyperbolic cluster states. We first review fault-tolerant protocols for hyperbolic surface codes in the circuit model, emphasizing Dehn-twist–based constructions on closed hyperbolic surfaces. We then summarize how fault tolerance is achieved in MBQC on the RHG lattice. Finally, we discuss how these frameworks suggest possible routes toward fault-tolerant logical operations on hyperbolic cluster states, leaving a complete analysis to future work.

\subsection{Fault-Tolerant Computation on Closed Hyperbolic Surface Codes}

In the circuit model, universality can be achieved by combining Clifford operations with magic-state injection and distillation \cite{bravyi2005universal, reichardt2006quantum}. For planar surface codes, lattice surgery provides a mechanism for performing joint logical Pauli measurements through merge-and-split operations between code patches \cite{horsman2012surface, litinski2019game}. These operations enable the implementation of entangling Clifford gates such as CNOT.

For closed hyperbolic surface codes, an analogous role is played by elements of the mapping class group of the surface. In particular, Dehn twists, homeomorphisms obtained by cutting the surface along a noncontractible cycle and regluing after a $2\pi$ twist, induce nontrivial symplectic transformations on homology classes and therefore act as logical Clifford operations \cite{breuckmann2017hyperbolic, Lavasani2019DehnTwists}. 

Importantly, Dehn twists on hyperbolic surface codes can be implemented using constant-depth local unitary circuits, up to a permutation of physical qubits that is local in the hyperbolic metric \cite{Lavasani2019DehnTwists}. The depth of the implementing circuit does not scale with the code distance, and the space overhead remains constant asymptotically. Throughout the protocol, the code remains an LDPC code with comparable distance, so standard decoding procedures remain applicable. These results demonstrate that hyperbolic surface codes admit fault-tolerant logical dynamics generated by topological automorphisms of the underlying surface.

\subsection{Fault-Tolerant MBQC on the RHG Lattice}

Fault-tolerant MBQC on the RHG lattice follows a different but closely related paradigm. The RHG lattice is a three-dimensional cluster state whose correlation structure is equivalent to a foliated surface code \cite{Raussendorf2006FaultTolerant}.
Logical Clifford operations are implemented through controlled manipulation of these topological features. In one formulation, logical operators are associated with correlation surfaces \cite{Raussendorf2007Topological, Tournaire2026dlatticedefect, bourassa2021blueprint}, and entangling gates are realized by braiding or deforming these correlation surfaces within the three-dimensional lattice. In another formulation, lattice surgery is interpreted within MBQC as performing joint logical Pauli measurements between encoded patches by appropriately tailoring the measurement pattern \cite{Horsman_2012_lattice_surgery, Herr2018LatticeSurgery}. These joint measurements enable the realization of entangling Clifford gates such as CNOT. In both formulations, universality is achieved by supplementing Clifford operations with magic-state injection.

\subsection{Perspectives for Hyperbolic Cluster States}

The foliated hyperbolic cluster states considered in this work, obtained by stacking $2z$ layers of periodic $\{8,3\}$ hyperbolic lattices and coupling them into a three-dimensional resource, occupy a conceptual middle ground between these two paradigms. On the one hand, each spatial layer inherits the structure of a hyperbolic QEC code, for which Dehn-twist–based logical operations are natural candidates. On the other hand, the full three-dimensional cluster state shares key features with the RHG lattice, suggesting that suitably designed measurement patterns along the foliated direction may realize joint logical measurements, and lattice-surgery–style operations.

The simplest gate is already implementable in the hyperbolic setup. Indeed, by the foliation principle \cite{Bolt2016FoliatedQECC}, the logical information encoded in the first primal layer is teleported to the last primal layer when measuring all individual qubits except in the final layer in $X$. This implements the logical identity gate from an input Hyperbolic code to an output one up to some logical Pauli operators determined by the measurement outcomes. Extending this mechanism to a universal set of fault-tolerant logical gates remains an open question: the precise extent to which Dehn-twist protocols for hyperbolic surface codes and RHG-style lattice surgery can be transferred to hyperbolic cluster states has yet to be established. One promising direction is to interpret the foliated $\{8,3\}$ construction as a hyperbolic analogue of the RHG lattice and to identify defect encodings whose transverse structure follows nontrivial cycles of the hyperbolic tiling. Another is to seek global measurement patterns that effectively implement Dehn twists of the underlying hyperbolic code within each layer, mediated by inter-layer entanglement, thereby realizing nontrivial logical transformations in an MBQC framework.

We therefore regard existing fault-tolerant schemes for hyperbolic QEC codes and for the RHG lattice as complementary templates rather than direct blueprints. 
A complete theory of fault-tolerant computation on hyperbolic cluster states will require specifying explicit logical encodings and measurement patterns that implement a universal set of logical operations within the hyperbolic cluster-state geometry. Establishing fault tolerance for such protocols will then entail identifying compatible decoding procedures and quantifying the resulting thresholds and resource overheads.

\section{Results and Discussion}

We have introduced a framework for constructing hyperbolic cluster states by foliating hyperbolic QEC codes based on periodic hyperbolic $\{p,q\}$ lattices. This construction preserves the topological fault-tolerance mechanism of the RHG cluster state, namely, the encoding of logical information in extended topological structures and the appearance of string-like fault propagation in the three-dimensional complex, while inheriting the constant encoding rate of hyperbolic QEC-code layers. In this sense, hyperbolic cluster states constitute a geometrically enriched generalization of the RHG resource, with the potential to reduce overhead at comparable fault-tolerance performance.

\begin{figure*}[t]
    \centering
    \begin{subfigure}[t]{0.48\linewidth}
        \centering
        \includegraphics[width=\linewidth]{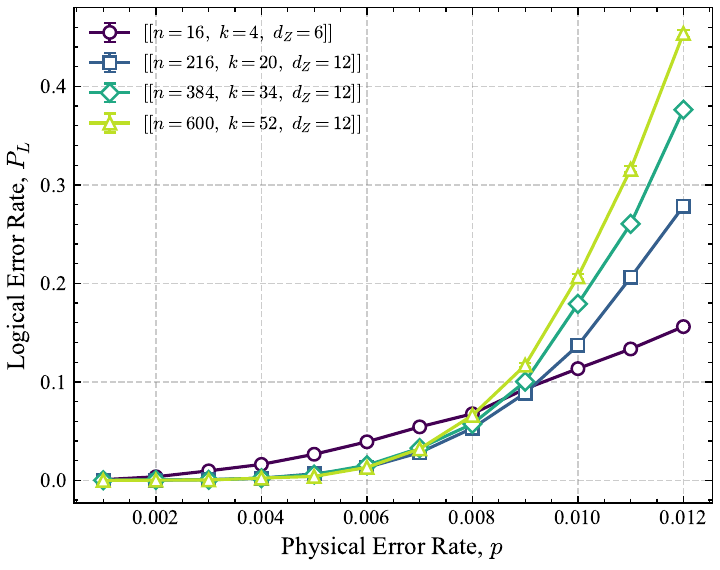}
    \caption{Logical $Z$ error rate as a function of the physical error probability under the circuit-level depolarizing noise model. The crossing of curves for increasing system sizes indicates a fault-tolerance threshold at $p_{\mathrm{th}}^{(Z)}\approx0.8\%$.}
    \end{subfigure}
    \hfill
    \begin{subfigure}[t]{0.48\linewidth}
        \centering
        \includegraphics[width=\linewidth]{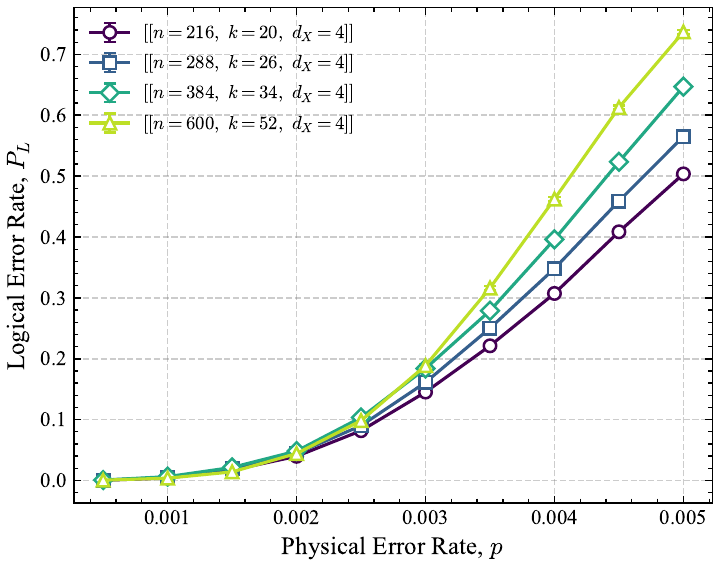}
        \caption{Logical $X$ error rate as a function of the physical error probability under the same circuit-level depolarizing noise model. The crossing of curves for increasing system sizes indicates a fault-tolerance threshold at $p_{\mathrm{th}}^{(X)}\approx0.25\%$.}
    \end{subfigure}
    \caption{Logical error rates for the $\{8,3\}$ hyperbolic cluster-state code under a circuit-level depolarizing noise model. Each data point is obtained from $2\times10^{4}$ Monte Carlo trials. Panels (a) and (b) show the logical $Z$ and logical $X$ error channels, respectively. Distinct thresholds are observed for the two channels.}
    \label{fig:thresholds}
\end{figure*}

To assess fault tolerance quantitatively, we performed memory experiments under the circuit-level depolarizing noise model described in Sec.~\ref{section:QEC on HCB}. For each physical error rate $p$, each data point in Fig.~\ref{fig:thresholds} is obtained via $2\times10^{4}$ Monte Carlo trials. Panels (a) and (b) report the logical $Z$ and logical $X$ error channels, respectively. For fixed foliation depth $2z=8$ in both channels, we observe the characteristic crossing of logical error-rate curves for increasing system sizes in each layer, indicating finite thresholds. For the $\{8,3\}$ hyperbolic cluster-state code studied here, we estimate thresholds of approximately
\[
p_{\mathrm{th}}^{(Z)} \approx 0.8\%, 
\qquad
p_{\mathrm{th}}^{(X)} \approx 0.25\%.
\]
These values are of the same order as thresholds ($0.75\%$) typically reported for Euclidean RHG-type constructions under comparable circuit-level noise assumptions, but the hyperbolic construction retains a constant encoding rate in the thermodynamic limit, see Eq.~\ref{eqn:encoding_rate} \cite{Raussendorf2007Topological}.

A pronounced asymmetry between the two logical channels is expected in the present $\{8,3\}$ implementation, since the underlying tiling is not self-dual. In particular, $Z$ errors are detected by $X$-type parity checks, whereas $X$ errors are detected by $Z$-type parity checks. For the $\{8,3\}$ hyperbolic cluster state, the $X$-type parity checks have weight $w_X = 5$, while the $Z$-type parity checks have weight $w_Z= 10$ (cf. Fig.~\ref{fig:parity-checks}). The lower weight of the $X$-type checks generally improves the detectability of $Z$-errors under circuit-level noise, consistent with the substantially higher observed threshold in the logical-$Z$ channel.

The primary limitation of the present numerical study is finite-size accessibility. A major computational bottleneck is the circuit-level fault characterization step: for each instance, we simulate all $N_F$ distinct single-location fault circuits to map each fault to its propagated syndrome and to construct the effective weighted decoding model used by MWPM. Together with the subsequent Monte Carlo sampling, this restricts us to comparatively small hyperbolic cluster states. The largest instance considered contains $n=600$ data qubits per sheet and $N_{\mathrm{total}}= 7{,}000$ total qubits in the 3D resource. Table~\ref{tab:overhead} summarizes the resource overhead of the $\{8,3\}$ hyperbolic cluster-state instances used in Fig.~\ref{fig:thresholds}.

In summary, this work establishes hyperbolic cluster states as viable fault-tolerant MBQC resources with thresholds comparable in scale to Euclidean RHG-type constructions, while unlocking constant-rate code families that can substantially reduce qubit overhead. More broadly, it initiates a geometry-driven program for MBQC in which curvature and topology become explicit design parameters for optimizing the underlying resource cluster states. We expect such insights to inform future MBQC architecture designs to move beyond Euclidean lattice constraints.

Several directions follow naturally from this work. Firstly, a systematic investigation of how code parameters and threshold performance depend on the underlying lattice geometry is still needed. Secondly, and most importantly, it remains to develop explicit fault-tolerant gate protocols on hyperbolic cluster states, including hyperbolic variants of defect braiding and lattice surgery, or measurement-based implementations of mapping-class-group operations such as Dehn twists. Together, these advances would turn hyperbolic cluster states from a fault-tolerant memory into a fully computational fault-tolerant architecture for MBQC with geometry-optimized overhead.

\begin{table*}[h!]
\centering
\begin{tabular}{ccccccccc}
\toprule
$n$ & $k$ & $d_Z$ & $d_X$ & $k/n$ & $N_{\mathrm{total}} =M_{\mathrm{HCS}}$ & $k/N_{\mathrm{total}}$ & $CZ_{\mathrm{HCS}}$ & $N_F$ \\
\midrule
16  & 4  & 6  & 2 & 0.250 & 196 & 2.04 $\times 10^{-3}$  & 272  & 4{,}276  \\
216 & 20 & 12 & 4 & 0.0926 & 2{,}520 & 7.937$\times 10^{-3}$ & 4{,}968 &  77{,}040 \\
288 & 26 & 12 & 4 & 0.0903 & 3{,}360 & 7.738$\times 10^{-3}$ & 6{,}624 &  102{,}720 \\
384 & 34 & 12 & 4 & 0.0885 & 4{,}480 & 7.589$\times 10^{-3}$ & 8{,}832 &  136{,}960 \\
600 & 52 & 12 & 4 & 0.0867  & 7{,}000 & 7.429$\times 10^{-3}$ & 13{,}800 &214{,}000 \\
\bottomrule
\end{tabular}
\caption{Parameters and resource counts for the $\{8,3\}$ hyperbolic cluster-state memory experiments with fixed foliation depth $2z=8$ (i.e., $z=4$ primal--dual layer pairs). The underlying hyperbolic CSS code places $n=E$ data qubits on lattice edges and encodes $k$ logical qubits, with distances $d_Z$ and $d_X$. The MBQC resource is characterized by the total number of cluster-state qubits $N_{\mathrm{total}}$ (equal to the number of single-qubit measurements $M_{\mathrm{HCS}}$), the total number of entangling gates $CZ_{\mathrm{HCS}}$, and the number of distinct single-location fault processes $N_F$ in the circuit-level depolarizing noise model.}
\label{tab:overhead}
\end{table*}

\section*{Acknowledgment}
We would like to thank the developers of the \texttt{hypertiling} package~\cite{schrauth2023hypertiling}, which was used to generate the lattice in Fig~\ref{fig:projection}. We acknowledge the use of Qiskit's AerSimulator for executing the MBQC circuits used in the simulations. SB and GT were supported by the NSERC Discovery Grant and the European Commission under the Grant \emph{Foundations of Quantum Computational Advantage} (both of SB).  
SR and AAM were supported by the NSERC Discovery Grant and PIMS Site Director Support Grant (both of SR). AAM was also supported in part by the Mitacs Globalink Research Award \emph{New Perspectives in Quantum Materials: Floquet Circuits and Magnetic Aspects of Hyperbolic Lattices} during the completion of the manuscript.

\section*{Code and Data Availability}
The data supporting the findings of this study, including the datasets used to generate Fig.~6 (threshold graphs), as well as the Python code used for simulation and analysis, are available from the authors upon reasonable request.

\bibliographystyle{unsrt}
\bibliography{references}
\end{document}